\documentclass[12pt]{article}

\usepackage{times}

\usepackage{graphicx}
\usepackage{epstopdf}
\usepackage{caption}
\usepackage{subfigure}
\usepackage{textcomp}

\newenvironment{sciabstract}{%
\begin{quote} \bf}
{\end{quote}}

\newcommand{\rpar}{\mathbf{r}}

\newcommand{\z}{z}                % superficie generica
\newcommand{\zsq}{\z_{\rm if}}    % superfice hielo/pelicula
\newcommand{\zqv}{\z_{\rm fv}}    % superficie pelicula/agua
\newcommand{\SqL}{i/f~}    % superficie pelicula/agua
\newcommand{\qLV}{f/v~}    % superficie pelicula/agua

\title{ 
\vspace*{-2cm}
{\bf  Title:} Surface phase transitions and crystal growth rates of ice
in the atmosphere \\ 
\vspace*{0.3cm}
\large
 {\bf Short Title:} Explaining the shape of snow crystals
}

\author
{ {\bf Authors:} Pablo Llombart,$^{1,2}$ Eva G. Noya,$^{1}$ and Luis G.~MacDowell$^{2\ast}$\\
\\
{\bf Affiliations:}
\normalsize{$^{1}$ Instituto de Qu\'{\i}mica F\'{\i}sica Rocasolano, Spain.}\\
\normalsize{$^{2}$Departamento de  Qu\'imica F\'isica, Universidad Complutense de Madrid, Spain.}\\
\\
\normalsize{$^\ast$Correspondence to:  lgmac@quim.ucm.es}
}

\date{}

\begin{document}


\baselineskip24pt


\maketitle


\begin{sciabstract}
   {\bf Abstract:}
With climate modeling predicting  a raise of at least 2 \textdegree{C} by
   year 2100, the fate of ice has become a serious concern, but we still do not
understand
how ice  grows (or melts). In the atmosphere, crystal growth rates of basal
and prismatic facets exhibit an enigmatic temperature dependence, and
crossover up to three times in a range between 0 and -40\textdegree. 
Here we use large scale computer simulations to
characterize the ice surface and identify a sequence of novel phase
transitions on the main facets of ice crystallites. Unexpectedly, we find that
as temperature is increased, the crystal surface  transforms from a disordered
phase with proliferation of steps, to a smooth  phase with small step density.
This causes the anomalous increase of step free energies
and provides the long sought explanation for the enigmatic crossover of snow
crystal growth rates found in the atmosphere.
\end{sciabstract}

%----------------------------------------------------------
{\bf One Sentence Summary:} Atomic scale description of the ice structure
predicts ice crystal growth rates and explains snow crystal shapes.

{\bf Note:} A revised version with title "Surface phase transitions
and crystal growth habits of ice in the atmosphere" accepted
the 6th march 2020 to appear in Science Advances (in press).

\newpage

{\bf Main Text:}

\section*{Introduction}

\begin{figure}
   \includegraphics[clip,scale=0.4]{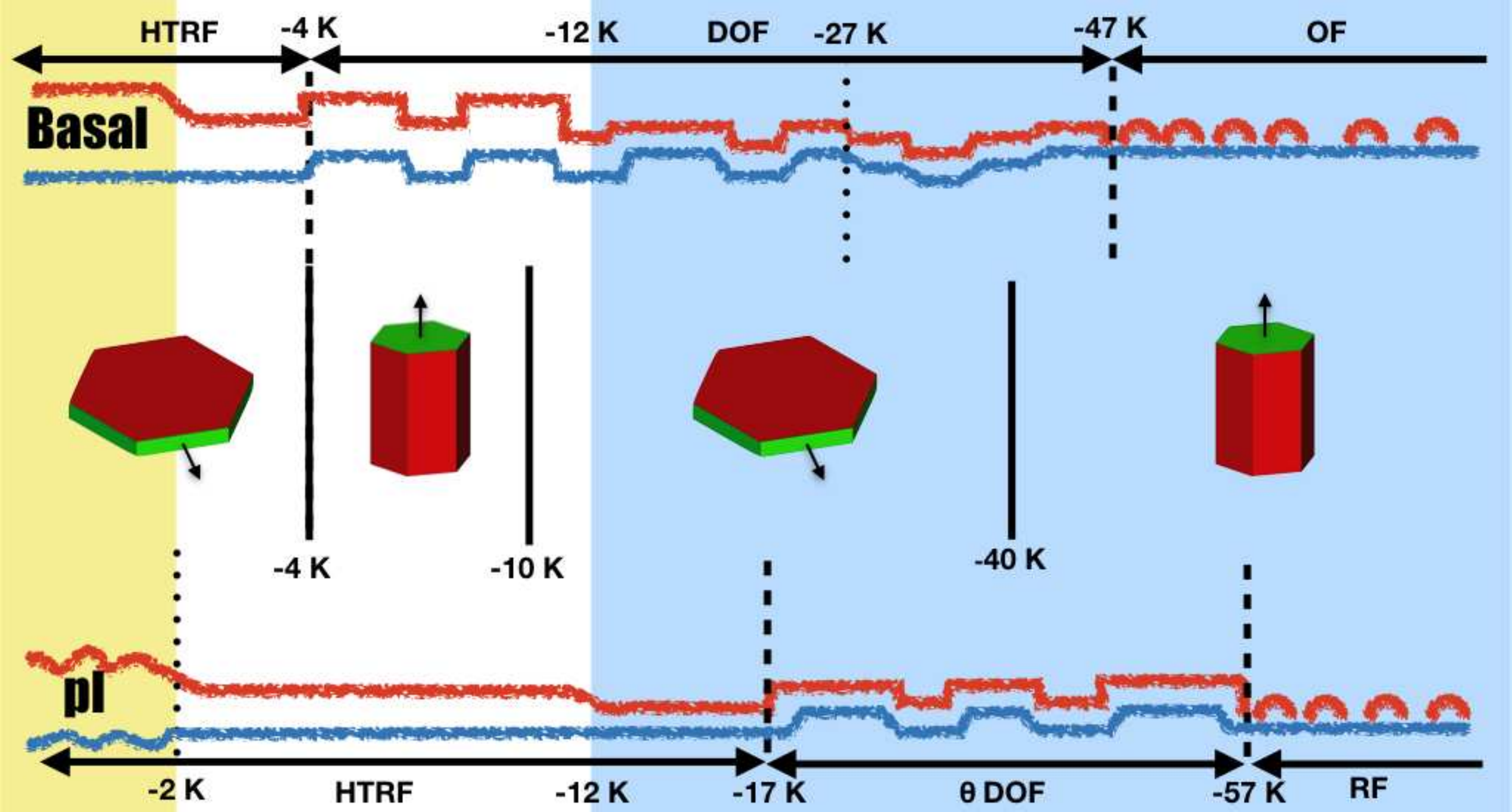}
   \caption{%\textsf
	{ 
          {%\scriptsize
           {\bf Growth of ice crystals at low supersaturation}. Middle panel: 
	     As temperature $\Delta T=T-T_t$ decreases below the triple point,
	     $T_t$, the habit of hexagonal ice prisms grown in the
         atmosphere change sharply from plate like to columnar at ca. -4~K, from
         columnar to plate like at -10~K and
         somewhat less sharply from plate-like to columnar at
	   -40~K \cite{bailey09}.
         The facet which grows faster
         as indicated by the arrows, dictates the prevalence of plates or
         columns. The change of crystal habits result
         from a crossover in the growth rates of the basal (red) and prismatic
         (green) facets. Top and bottom:   Sketch  of surface structural
         evolution  with temperature as found in our work. Blue lines represent the \SqL surface, and
         red lines represent the \qLV surface. The basal surface (top row) is a
         High Temperature Reconstructed Flat phase from $\Delta T=0$ to -4~K. It
          becomes a Disordered Flat Phase in the range between ca. -4~K to
          -47~K and is transformed into an  Ordered Flat phase at lower
          temperatures. In this phase, surface disorder resulting form patches of liquid-like
          molecules remains. The prismatic surface (bottom row)  is a High
          Temperature Reconstructed Flat phase all the way from 0~K to -17~K, but
          is very close to the roughening transition at $\Delta T>-2$~K in our
          model. In the range from -17~K to -57~K it is a Disordered Flat Phase and becomes an
          Ordered Flat phase below -57~K. At the transition from DOF like to
          HT-RF phases,  step free energies increase anomalously and result in the
          crossover of crystal growth rates. The shaded areas illustrate the
          temperature range where melting of full bilayers has been
	    accomplished.
          Blue, less  than one full bilayer; white, less than two full bilayers;
          yellow, more than two full bilayers.
          \label{fig:sketch}
 }}}
\end{figure}

The Nakaya diagram  documents the hidden mystery of snow
crystal growth \cite{nakaya54,vandenheuvel59}: as temperature is cooled down 
from 0 to -40\textdegree{C}, ice crystals in the atmosphere
change their habit from plates, to columns, to plates
and back to columns in a puzzling and unexplained
sequence that holds the key to our understanding of
the ice surface (Fig. \ref{fig:sketch}).

After heterogeneous nucleation \cite{kiselev16,qiu17}, there still is a long way
before ice crystallites adopt the micron size found in cirrus clouds
\cite{peter06}. Whether the ice embryos transform into mature columnar or 
plate like hexagonal prisms depends on the relative growth rate of basal and 
prismatic facets \cite{sei89,demange17}. But what drives the crossover
of basal and prismatic growth rates, and how such changes are related
to the ice surface structure remains completely unknown to date
\cite{furukawa07}.  Kuroda and Lacmann speculated 
that the occurrence
of surface phase transitions  could result in sudden
changes of the crystal growth rate, and provide an explanation for
the observed crystal habits \cite{kuroda82},  but unfortunately the  
experimental verification 
of this hypothesis has remained elusive to 
date \cite{wei01,bluhm02,smit17,slater19} and our current understanding is 
still far from providing a molecular explanation 
\cite{bartels-rausch13,ball16,libbrecht17,slater19}. However, recent findings 
point to
the significance of the heterogeneous in plane structure of ice's premelting
film \cite{qiu18}.

Here we show that, similar to under-cooled bulk water, 
the facets of ice  exhibit
a different phases as temperature is cooled down below
0~\textdegree{C}. Particularly, we identify a  transition from
an Ordered Flat to a  Disordered Flat Phase that is
akin to   microphase separation of terraces on the crystal  surface
 \cite{jagla99,rommelse87}.
Upon increasing temperature, the premelting film thickness grows
and the surface structure flattens again, resulting in the
unexpected increase of nucleation step free energies and
the crossover of crystal growth rates exactly as found in laboratory
and atmospheric field studies \cite{libbrecht17,bailey09}.

\section*{Results}

\begin{figure}
%   \vspace*{-1.5cm}
   \includegraphics[clip,scale=0.15]{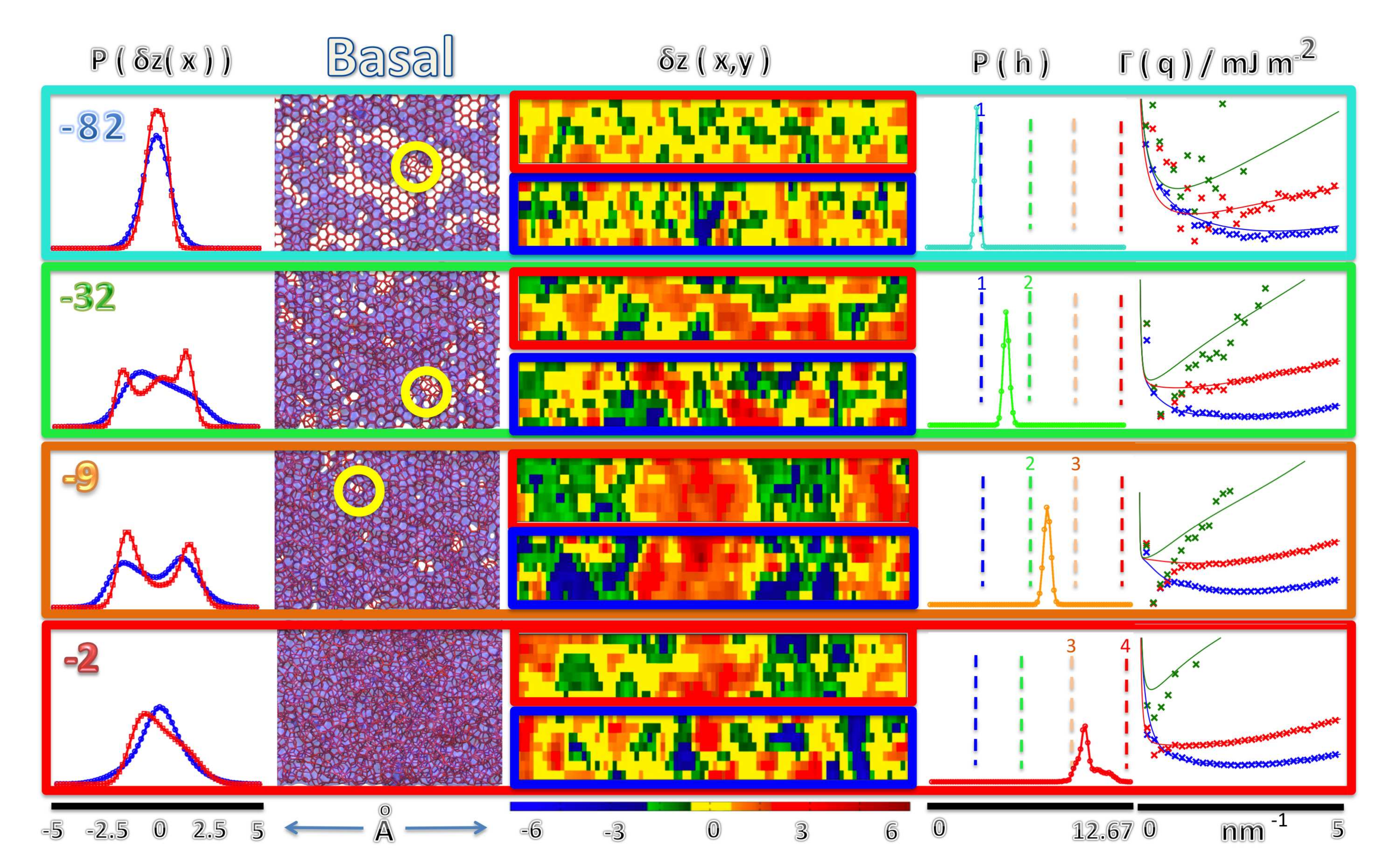}
   \caption{
	  {%\scriptsize
	%\textsf
	{ 
	{\bf Surface fluctuations on the basal facet}. a) Probability distribution
	of \SqL (blue) and \qLV (red) surface fluctuations, as measured by the
	deviations of the interface position $\z_{\alpha}(x)$ about the average 
	surface $\overline z_{\alpha}$, for $\alpha=\{if,fv\}$. 
	Results are shown for different temperatures as indicated in the legend. 
   b) Snapshots of the basal surface at the same four temperatures.
   Red lines show the connected hydrogen bond network of all solid-like and liquid-like
   water molecules.  The violet patches represent disordered liquid-like molecules.  
   At low temperature the surface is mainly a regular hexagonal honeycomb, with a few patches 
   of liquid-like molecules siting on interstitial positions (as indicated by
   the yellow circles). The
   extent of filled interstitial positions increases as the premelting layer
extends on the surface.  c) Plot showing a snapshot of
 local surface height fluctuations 
$\delta \zsq(\rpar)$ (bottom, blue
frame) and $\delta \zqv(\rpar)$ (top, red frame) on the basal ice face.
Notice the emergence of large scale correlated patches for the DOF
phase in the temperature range $\Delta T=-32$ to -9~K (See movie M1 and M2) 
%in Extended data section. 
The patches disappear at high
temperature as the surface flattens again. d) Distribution of average film
thickness, $h$, as temperature increases. Dashed vertical lines represent full
layers in units of the molecular diameter. d) Wave-vector dependent stiffness
coefficients, as obtained from the inverse surface structure factor
for \SqL correlations (blue), \qLV correlations (red) and crossed \SqL-\qLV
correlations (green). Crosses are results from simulation, full lines are fit to
the SG+CW model. The results show that all surfaces are smooth, as indicated by the $q\to 0$ divergence
of $\Gamma(q_x)$. Notice the sharp minimum appearing at intermediate
length scales in the temperature range $\Delta T=-32$~K to $-9$~K where the DOF
phase is present.
	  \label{fig:basal}
 }}}
\end{figure}

We perform large scale computer simulations of the TIP4P/Ice point
charge model of water \cite{abascal05} on elongated rectangular surfaces meant to
characterize large wave-length correlations along one direction.  A bulk ice sample is 
placed in vacuum at temperatures below the model's triple point, $T_t=272$~K. After a few
nanoseconds, the surface spontaneously develops a layer of quasi-liquid
disordered molecules that can be readily distinguished from the underlying
bulk crystal network (Fig.\ref{fig:basal}-b). 
Averaging the
position of the outermost solid and liquid-like atoms of the premelting layer
about points $\rpar$ on the plane of the interface,
we are able to identify  distinct ice/film, $\zsq(\rpar)$  and film/vapor
$\zqv(\rpar)$ surfaces \cite{benet16}, 
which separate the premelting film from the bulk solid
and vapor, respectively (see Methods and %Extended Data 
Fig. S1)
The average thickness of this layer
grows from about  3{~\AA} at $\Delta T = T - T_t=-82$~K to 9{~\AA} at
$\Delta T=-2$~K with little measurable anisotropy (%Extended Data 
   Fig. S2) \cite{conde08}.
However, as recently observed for the basal plane in the mW
model \cite{qiu18}, this 
thin disordered layer is laterally inhomogeneous up to about
$\Delta T=-9$~K (Fig. \ref{fig:basal}-b and %extended
S3). Our study reveals that the heterogeneity is also found 
to a similar
extent on the prismatic plane, with little qualitative differences (Fig.
   \ref{fig:pI}-b and %Extended Data 
Fig.S4). 
   We quantify the $\alpha=\{if,fv\}$ surface fluctuations along the long direction, $x$,
by studying deviations of the local surface position, $\delta \z_{\alpha}(x)$, 
about the average surface, $\delta \z_{\alpha}(x) = \z_{\alpha}(x) -
\overline\z_{\alpha}$ (methods). This analysis is essential to reveal differences
between basal and prismatic planes 
and allows us to identify a number of phase transitions along the 
sublimation line which cannot possibly be inferred from visual inspection of 
the snapshots (Fig \ref{fig:basal}-a, \ref{fig:pI}-a and %Extended Data 
Fig.S5, Fig.S6).

%\vspace*{-1cm}

\subsection*{Structure of the basal facet}
At low temperature ($\Delta T=-82$~K), the basal plane consists of a relatively
ordered flat solid surface (OF), as revealed by a singly peaked, close to Gaussian
distribution of both \SqL and \qLV surface fluctuations (Fig.\ref{fig:basal}-a
and %Extended Data 
Fig.S5-a). From the
snapshots (Fig.\ref{fig:basal}-b), the solid surface is formed mainly of an oxygen-unreconstructed
stack of chair hexagons \cite{fletcher92,buch08,pan10}. Patches of disordered
liquid-like molecules are found, and often show a tendency to sit on
interstitial positions  at the center of the primary hexagonal mesh (as in
the so called Honeycomb Fletcher phase) \cite{fletcher92}.
The distribution remains uni-modal up to $\Delta T=-52$~K, but somewhat 
broadens,
revealing a large increase of disorder in this temperature
interval which is consistent with observations of Sum Frequency Generation
Spectroscopy \cite{wei01}.  At  $\Delta T=-42$~K, however, the distribution of
$\delta \zqv(x)$ develops a distinct trimodal character (Fig. 
S5-a). 
A main peak is centered at the mean surface position, and two other peaks
appear to the left and right. 
The central peak of the trimodal distribution for $\delta \zqv(x)$ gradually 
fades away into a bimodal, which persists up to $\Delta T=-6$~K. In a narrow
range between -9~K to -6~K, the distributions of $\delta \zqv(x)$ and
$\delta \zsq(x)$ are both fully bimodal and congruent.
Finally, at the temperature of -2~K,
the bimodal collapses sharply into one single uni-modal distribution.

From our analysis, the outer \qLV surface of the premelting film exhibits
a bimodal distribution centered at the mean surface location all the way from
-22~K to -6~K. The onset of bimodality very much correlates with the vanishing
of the (ppp-polarized) dangling OH bond stretch
observed in SFG experiments \cite{wei01}. 
The separation between peaks  in the bimodal
is approximately 3.1~\AA , somewhat smaller than the expected bilayer distance 
of $c/2=3.65$~\AA. 
Furthermore, the bimodal evolves after the appearance at
low temperature of a trimodal distribution with a main peak centered at the 
mean surface position,
as if ice melting resulted in the formation of water-like molecules at
half integer lattice positions (Extended Data 
Fig.S5-a). %\ref{fig:ext_datas_basal}-a) %,b). 

\subsection*{Preroughening and smoothening transitions}

A strongly disordered  phase
consisting of a smooth surface  with large scale
step proliferation has been documented in the literature for Solid on Solid 
models (SOS) and is known as a disordered flat phase (DOF) \cite{rommelse87}. 
Exactly as suggested by the extended SOS model, we
observe that the low temperature smooth flat phase transforms into a locally 
rough interface at a  preroughening transition, $\Delta T_{pr}\approx -47$~K
\cite{rommelse87}. 
However, contrary to the usual scenario, at high 
temperature the DOF phase does not undergo roughening, but rather,
transforms across an apparently first order surface phase transition into
a new flat phase at a   {\em smoothening temperature}, $\Delta T_s\approx -4$~K. This
transition has not been previously reported but 
is consistent with existing surface phase diagrams for the extended SOS
model \cite{prestipino95,weichman96,bastiaansen96}. Following the notation of
SOS literature, we call this a High Temperature Reconstructed Flat phase
(HT-RF).

\begin{figure}
   \includegraphics[clip,scale=0.16]{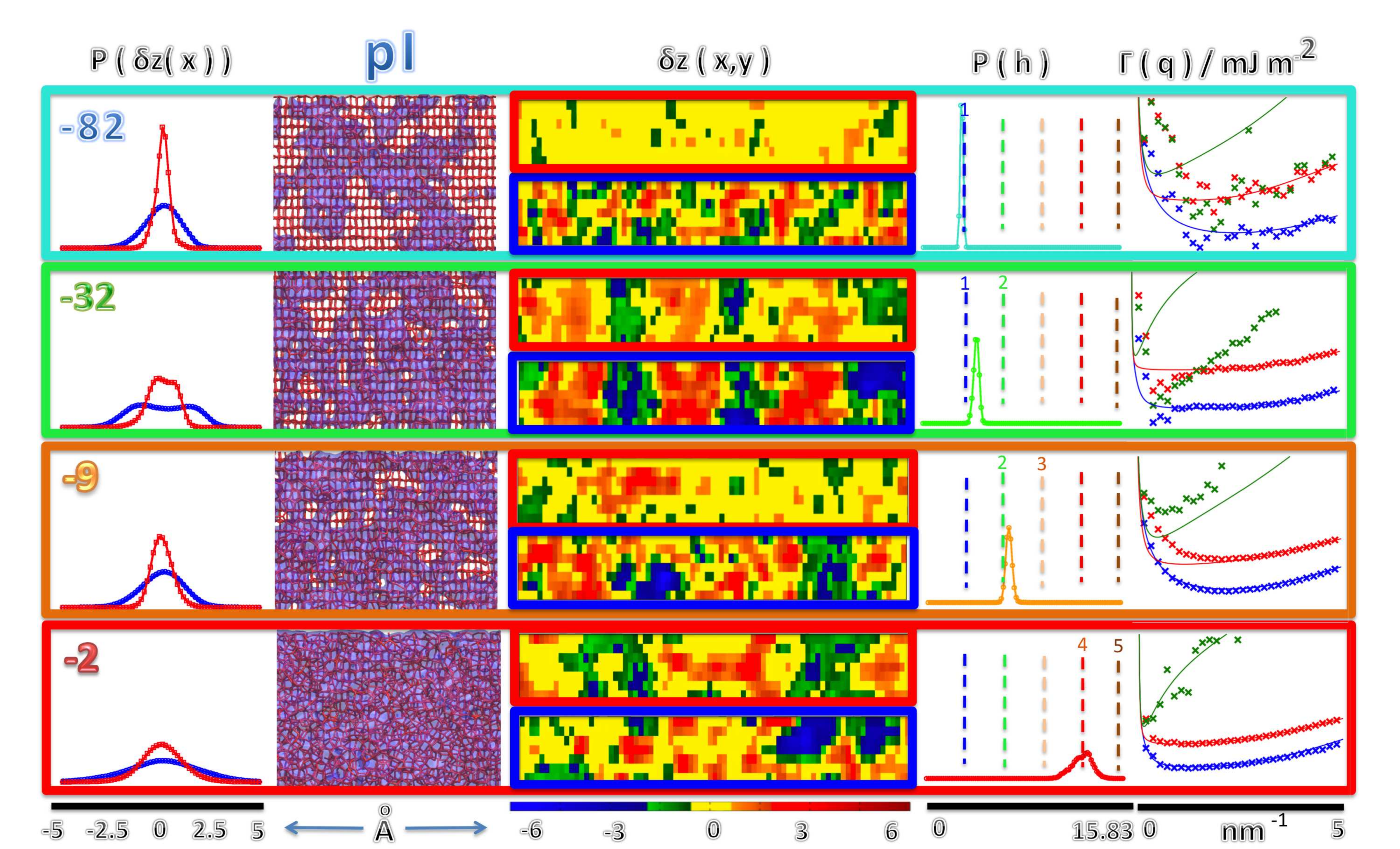}
   \caption{%\textsf
	{ 
	  {%\footnotesize
	{\bf Surface fluctuations on the prismatic facet}. Content displayed
as in Fig.\ref{fig:basal}. a) A bimodal distribution in this facet is only
visible at temperature $\Delta T=-32$~K.  b) Here the snapshots show the characteristic
rectangular mesh of the prismatic facet. At low temperature  the liquid-like molecules form 
patches on the solid surface as in the basal face.  c) Emergence of large correlated domains signal a
DOF phase that is clearly visible at $\Delta T=-32$~K and vanishes  at significantly lower 
temperatures than for the basal facet (see movies M4 and M5 in Supplementary
Materials Section)
%in Extended data section).
d) Notice the transition of premelting layer thickness across integer multiples of the molecular diameter
occurs for the prismatic facet at the same temperature as for the basal facet.
e) Likewise, the sharp minimum of the stiffness coefficient is visible only at
and below $\Delta T=-32$~K. 
 \label{fig:pI}
 }}}
\end{figure}

As expected for the preroughening scenario, the transition to a DOF phase
is strongly correlated with the growth of a premelting film in a loosely
layer wise fashion \cite{weichman96,jagla99}. Clearly, we observe that the stabilization of the
bimodal DOF phase at $\Delta T=-22$~K results after the full formation of a 
second layer of premelted ice, as revealed by the mean location of
premelting layer fluctuations in Fig.\ref{fig:basal}-d  
(%Extended Data 
Fig. S5-b). The transition from
DOF into the HT-RF phase is also accompanied by the formation of
a full third premelted layer, as revealed by the shift in the
premelting layer thickness from $\Delta T=-9$ to -2~K (Fig.\ref{fig:basal}-d). 

The DOF phase is not only characterized by step proliferation and strong local
disorder. Also, the steps are highly correlated, and depending on the nature
of the DOF phase, can exhibit diverging (finite) parallel correlations at a
second (first)
order preroughening transition \cite{bastiaansen96}.
We explore the extent of parallel correlations  with a 
plot of the local surface height $\delta \zsq(\rpar)$ and $\delta \zqv(\rpar)$
fluctuations
(Fig.\ref{fig:basal}-c), %Extended Data 
Fig. S7-a). At low
temperature, the \SqL surface exhibits  small amplitude up and  
down domains, with small correlation lengths. At $\Delta T=-9$~K, where the
DOF phase is fully formed, we observe very large up and down domains of
about 9~nm in length, that
remain correlated over the full simulation box, as is visible both
in the figure and in the accompanying  movie (%Extended Data, 
(M1 and M2).
Our results are consistent with glancing angle x-ray experiments which
reported the appearance of a large surface correlation length in the
nanometer range, and the sharp disappearance of the long range correlations
close to the triple point \cite{dosch95}.

The nature of these correlations can be quantified from the 
wave-vector dependent surface
structure factor. Plots of the related effective stiffness,
$\Gamma_{\alpha\beta}(q_x) 
= \frac{k_BT}{A\langle \z_{\alpha}(q_x)\z_{\beta}(q_x)\rangle}$ are shown in
Fig.\ref{fig:basal}-e (%Extended Data 
Fig. S5-c). The 
results confirm that both at the preroughening
and at the smoothening transition, the parallel correlations 
remain finite at
$q_x\to 0$, as can be inferred from the strong divergence of the
effective stiffness coefficients. Everywhere in the region where a DOF 
phase is present, however, the strongly correlated up and down
domains are detectable as a sharp and deep minimum of the stiffness
coefficients. 
A full theoretical description of 
this strong enhancement  of large but finite correlations seems difficult.
, which we attribute to non-Gaussian fluctuations of the correlated step heights. 
However,  we can
definitively observe in our results how the location of the sharp
minima at intermediate wave-vectors decreases
as the size of the correlated domains in Fig.\ref{fig:basal}-c increase
%(Extended Data, Fig \ref{fig:hxy_basal}-a).

\subsection*{Structure of the Prismatic Facet}

The study of surface fluctuations on the prismatic facet is significantly
different. A low temperature flat phase, which preserves
the expected low temperature rectangular mesh  is observed
at $\Delta T=-82$~K (Fig.\ref{fig:pI}-a and b).  
At this temperature, the distribution
of both $\zsq(x)$ and $\zqv(x)$ are unimodal 
but become
gradually broader and skewed to the left as temperature is increased
%(Fig.\ref{fig:pI}-a). 
At $\Delta T=-32$~K, however, the distribution of $\zsq(x)$ and $\zqv(x)$ 
become slightly bimodal.
From $\Delta T=-12$~K
to $\Delta T=-2$~K, the distribution becomes again completely unimodal and
Gaussian like, but has broadened abruptly at $\Delta T=-2$~K, signaling the
approach of a roughening transition \cite{weeks80}. Although the order parameter
is not sharp enough to reveal a strong bimodality, there appear a
large number of steps in the range between -62~K to -22~K with
distributions that seem to resemble a $\theta$DOF phase (similar to
   the DOF phase, but with a continuous change of the step
coverage \cite{weichman96}.)  The transition from a DOF phase to a new high temperature 
reconstructed flat phase (HT-RF) is visible in the
surface maps (Fig \ref{fig:pI}-c), where large scale domains appear to end at
-32~K. This is also visible in the surface structure factors depicted
in Fig.\ref{fig:pI}-e, which reveal again diverging stiffness coefficients
at $q_x\to 0$ (whence, flat phase), and the complete disappearance of the sharp minimum
beyond this temperature (whence, loss of long range surface order).  
The transformation of
the surface structure across the $\theta$DOF phase is also accompanied with 
continuous increase
of the premelting film (Fig.\ref{fig:pI}-d), but the transition from two to three layers and beyond
appear to occur almost at the same temperature as in the basal facet. 
On the contrary, the transition from the $\theta$DOF phase to the HT-RF phases is
distinctly different at the basal (ca. $T_s = -4$~K) and the prismatic
(ca. $T_s = -27$~K) facets. Interestingly, these two temperatures, together 
with 
the full completion of a second bilayer at $T_l=-12$~K are in very close
agreement with the crystal habit changes observed in the Nakaya diagram.

\begin{figure}
   \includegraphics[clip,scale=0.16]{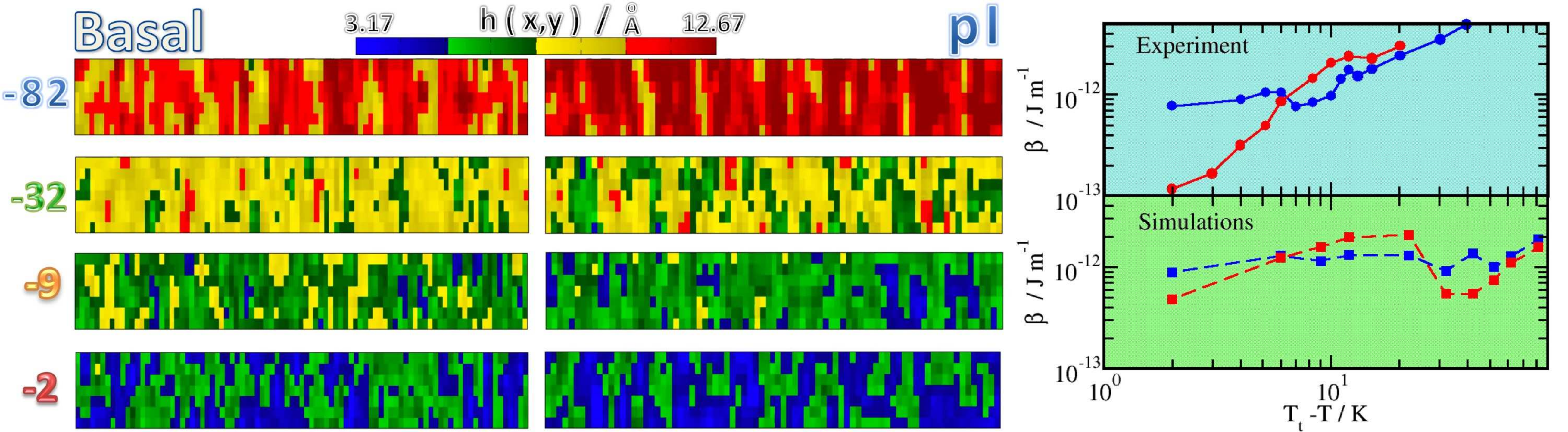}
	  {\footnotesize
   \caption{%\footnotesize \textsf
	{ 
	   {\bf Fluctuations of premelting thickness and step free energies}
   Figures a) and b) show a surface plot of instantaneous premelting thickness,
   $h(\rpar)$ for basal (a) and prismatic (b) surfaces (see movie M3 and M6 in 
   Supplementary Materials section)
   %Extended data section). Temperatures are indicated. 
   Notice the absence of large correlated domains at all
   temperatures, in marked contrast with the $\delta \zsq(\rpar)$ and $\delta
   \zqv(\rpar)$ surfaces shown in Fig.\ref{fig:basal} and Fig.\ref{fig:pI}.
c) Step free energies as obtained from a fit of the mean field
	   SG+CW model to the regular (Gaussian) part of the stiffness
	   coefficients in Fig.\ref{fig:basal}-e,\ref{fig:pI}-e for
	   the basal (blue) and prismatic (red) facets. Results (squares
	   with dashed lines)
	   are compared to experimental data 
	   (circles with full lines) \cite{libbrecht17} and displayed
	   on two different figures to avoid crowding.
	   The step free energies from the fit exhibit a crossover at 
	   ca.~-6~K, and overall a very similar trend as compared to experiment.
 \label{fig:beta}
 }}}
\end{figure}

We find that on average, both for the basal and the prismatic facets, $\zqv(\rpar)$
follows broadly the corrugation imposed by the $\zsq(\rpar)$, and the
fluctuations of premelted film thickness  appear as a broad unimodal
distribution with no sign of bimodality or diverging correlation lengths
(Fig.\ref{fig:basal}-d and Fig.\ref{fig:pI}-d). Accordingly,
there are apparently no layering transitions along the sublimation line
in the temperature range studied, 
at least
in the thermodynamic sense. The lack of large correlation lengths of
bimodality  is very clearly observed in 
the surface plots of local film thickness  $h(\rpar) = \zqv(\rpar) -
\zsq(\rpar)$ (Fig.\ref{fig:beta}-a and movies M3 and M6) 
and is consistent
with SFG experiments and preliminary indications for the mW model (%Extended Data
Fig.S7-c and S8-c) \cite{smit17,qiu18}.

\subsection*{Step Free Energies}

The significance of DOF phases has been well 
documented \cite{rommelse87,prestipino95,jagla99}. The step
proliferation is akin to a strong reduction of the step free energy,
and the sharp decrease of the threshold for linear
growth \cite{prestipino95}.
It is therefore expected that, as temperature raises
across the smoothening transition from a DOF to a HT-RF phase, 
the crystal growth rates will decrease anomalously.

We provide a quantitative test of this expectation using 
a model of coupled capillary wave and sine Gordon Hamiltonians
for the spectrum of surface fluctuations
discussed recently (SG-CW) \cite{benet16}. The capillary wave Hamiltonian describes
the \qLV surface fluctuations, while the likelihood of step proliferation
in the underlying \SqL surface is described with a sine Gordon Hamiltonian.
Both models are coupled with an interface potential that sets the
equilibrium film thickness as the difference between $\zqv(\rpar)$ and 
$\zsq(\rpar)$.  
The model can be fit to the regular (Gaussian) part of the
surface stiffness, 
(Fig.\ref{fig:basal}-e,\ref{fig:pI}-e) and provides phenomenological coefficients
for the surface tension and the step free energy, $\beta$ (Methods). 
The non-monotonous behavior observed for $\beta$ correlates with
the behavior measured experimentally for terrace spreading rates
\cite{sei89}, and is qualitatively very similar
to step free energies  obtained from 
growth measurements on snow crystals (Fig \ref{fig:beta}-b) \cite{libbrecht17}.
Most importantly, our results confirm the suggested scenario of
anomalous increase of $\beta$ exactly at the smoothening transitions
of basal and prismatic facets, respectively. 

\section*{Discussion}

Ice crystallites in the atmosphere are expected to grow by a two dimensional
nucleation process. The crossover from  column  to  plate like crystal
habits that is observed experimentally can only proceed by a corresponding 
crossover in the relative growth rates of basal and prismatic facets. 
In our work, we show that in the range of about 80~K below the melting point,
the main facets of ice may be found in
up to three different surface phases with varying degree of surface disorder,
as postulated by Kuroda and Lacmann with amazing intuition almost forty
years ago \cite{kuroda82}. 
The accompanying phase transitions provide the mechanism for
a non-monotonous change of the relative step free energies for two dimensional
nucleation. Most notably, we observe a premelting mediated process of
surface smoothening, which results in the anomalous increase of
step free energies. 
This results in the crossover of relative
growth rates of basal and prismatic facets that is required to explain
the Nakaya diagram. 

The explanation of the long standing problem of snow crystal
shapes (Fig.\ref{fig:sketch}) proofs that we have now a close to
complete molecular
description of the surface structure and crystal
growth rates of ice in the atmosphere, with immediate implications
for atmospheric sciences, glaciology and climate modeling.

% In setting up this template for *Science* papers, we've used both
% the \section* command and the \paragraph* command for topical
% divisions.  Which you use will of course depend on the type of paper
% you're writing.  Review Articles tend to have displayed headings, for
% which \section* is more appropriate; Research Articles, when they have
% formal topical divisions at all, tend to signal them with bold text
% that runs into the paragraph, for which \paragraph* is the right
% choice.  Either way, use the asterisk (*) modifier, as shown, to

\section*{Materials and Methods}

\subsection*{Force field}

Our study is performed with the TIP4P/Ice model of 
water \cite{abascal05}. This model was purposely designed
to best describe the properties of ice. It predicts a
melting point of T=272~K, in excellent agreement with
experiment, and reproduces the most relevant surface
properties at this temperature, such as liquid-vapor surface tension ($\gamma_{lv}=82$~mN/m
calculated by ourselves, compared with $\gamma_{lv}=75.7$~mN/m from experiment),
and solid-liquid surface tension ($\gamma_{sl}=29.8$~mN/m
from Ref \cite{espinosa16}. compared with recommended results $\gamma_{sl}=28$~mN/m 
by Pruppacher and Klett \cite{pruppacher10}).  The precise location
of the surface phase transitions observed here could somewhat change with
the molecular model employed, but we expect the generic features to
be quite generally observed for other accurate intermolecular potentials.

\subsection*{Initial configurations}

Initial configurations
are prepared from a perfect unit cell in pseudo-orthorhombic
arrangement, consisting of two layers of hexagonal rings perpendicular
to the hexagonal ${\bf c}$ axis and a total of 16 water molecules. 
For the basal interface, we arrange a stack of $46\times 8\times 8$  cells of
47,104 molecules, with
the long direction, $x$, aligned along the ${\bf b}$  axis of the
pseudo-orthorhombic cell, corresponding to the so called (basal)[pII] surface
arrangement described by Davidchack and used in our previous 
work \cite{davidchack06,benet16,benet19}.

For the pI interface, the simulation box is prepared
from a stack of $40\times 8\times 8$ unit cells of N=40,960 molecules, 
with the long direction, $x$, aligned along the ${\bf a}$ axis. This
corresponds to the (basal)[pII] arrangement in our recent work.
For each such arrangement, we prepare an independent hydrogen bond
network as described in Ref. \cite{buch98}. After forming the ice slab,
we perform NpT simulations  of the bulk solid at the desired temperature
in order to obtain equilibrated unit cell dimensions. The solid is
then scaled to the average equilibrium cell value, placed in
vacuum, and equilibrated again in the NVT ensemble under periodic boundary
conditions.

\subsection*{Computation details}

Large scale simulations are carried out on the Mare Nostrum  IV
facility at  Barcelona Supercomputing center from the
Spanish National Supercomputing Network. Classical Molecular Dynamics simulations
are performed with the GROMACS package \cite{gromacs4}. Trajectories are evolved with the
velocity-verlet algorithm. Both the Lennard-Jones and Coulomb interactions are truncated
at 0.9~nm. The electrostatic interactions are calculated using
the Particle Mesh Ewald method. Simulations are thermostated with the velocity rescale
algorithm \cite{bussi07}, and the Berendsen barostat when required. A relaxation
time of 2~ps is used for both the thermostat and barostat.  The timestep employed is 0.003~ps. 
The simulations proceed over 0.9~$\mu$s, with  a long equilibration time of 225~ns  and 675~ns 
for the production stage.

\subsection*{Surface analysis}

Prior to determining the \SqL and \qLV surfaces, we label water
molecules as either solid or liquid-like, using the $\bar q_6$
parameter \cite{lechner08}. Water like molecules are those with a $\bar q_6$
parameter below a threshold $\bar q_6^*(T)$. In order to determine the
threshold, we simulate the probability distribution of $\bar q_6$ at a number
of temperatures in either bulk solid or liquid water. The threshold value
$\bar q_6^*(T)$ is determined such that the number of mislabeled liquid molecules
on the solid phase is equal to the number of mislabeled solid molecules on
the liquid phase.  A cluster analysis is performed to determine which molecules pertain to 
the condensed phase.  Water molecules with oxygen atoms at a distance less than 3.5~{\AA} belong 
to the same cluster.  The \SqL surface is determined from the positions of solid-like atoms in the
largest solid cluster.  We use the heights of  of the four topmost (or bottommost)
solid atoms of the upper (lower) interface. At a given point $\rpar$ on
the plane of the interface, we find all the solid like atoms lying within
a rectangular prism centered at $\rpar$. The base of the prism is taken to
be that of the pseudo-orthorhombic unit cell of corresponding orientation.
The surface height $\zsq(\rpar)$ at that point is determined from the average
location of the four uppermost solid like atoms. At the same point, the liquid
surface for the upper (lower) interface is determined by averaging the position of the 
uppermost (bottom-most) for liquid-like molecules of the cluster of condensed
molecules lying within a rectangular area of $3\sigma \times 3\sigma$ Lennard-Jones
molecular diameters. The surfaces $\zsq(\rpar)$ and $\zqv(\rpar)$
are determined over points on a grid on the plane of the interface. The grid
has twice as many points as unit cells along the $x$ direction, and just as many
points as unit cells along the $y$ direction. We perform the surface analysis
from the set $\{\zsq(\rpar)\}$ and $\{\zqv(\rpar)\}$ of points on the
grid \cite{benet16}.
The instantaneous mean position of the \SqL surface, $\overline \zsq$ is determined as
the lateral average of $\{\zsq(\rpar)\}$ over all points on the grid. From this
value, we obtain $\delta \zsq(\rpar) = \zsq(\rpar) - \overline \zsq$. 
The laterally averaged fluctuations $\delta \zsq(x)$ are obtained from
$\delta \zsq(\rpar)$  upon averaging along points $y$.  $\delta \zqv(\rpar)$ and 
$\delta \zqv(x)$ are   obtained likewise.  Fourier transforms of $\delta \zsq(x)$ and 
$\delta \zqv(x)$ are obtained by summing $\delta \z_{\alpha}(x) e^{-i q_x x}$ over 
all points along $x$.  Instantaneous local film heights
are obtained as $h(\rpar) = \zsq(\rpar) - \zqv(\rpar)$. The instantaneous
average film thickness is obtained from the mean of $h(\rpar)$ over the points
of the grid.

\subsection*{SG-CW model and fit}

We describe the coupled \SqL and \qLV surface fluctuations with
an extended Sine Gordon model for the \SqL surface and the Capillary Wave
model for the \qLV surface \cite{weeks80}. The two terms are coupled via
the interface potential, $g(h)$, with the premelting film thickness
given by $h(\rpar)=\zqv(\rpar)-\zsq(\rpar)$. The full Hamiltonian
is given by: 
\begin{equation}\label{eq:sgcw}
        H=\int d\rpar\left[\frac{\tilde\gamma_{iw}}{2}(\nabla \zsq)^2 + 
	     \frac{\gamma_{wv}}{2}(\nabla \zqv)^2 + 
	    % \frac{\gamma_{iv}}{2}\nabla \nabla \zsq\cdot \nabla \zqv 
	    \gamma_{iv}\nabla \zsq\cdot \nabla \zqv 
	    - u\cos(q_z\zsq) + g(\zqv-\zsq)
	     \right ]
\end{equation}
where $\tilde\gamma_{iw}$ is the bare stiffness coefficient, $\gamma_{wv}$ is the
water-vapor surface tension, $\gamma_{iv}$ dictates the coupling of surface
deformations, $u$ accounts for the cost of moving the surface
$\zsq$ away from integer lattice spacing, $g(h)$ is the interface potential
dictating the equilibrium film thickness, and 
$q_z$ is the wave-vector for a
wave-length equal to the lattice spacing. This Hamiltonian can be 
expanded to quadratic order in deviations away from the mean surface
positions $\overline \zsq$ and $\overline \zqv$, and yields for the
thermally averaged surface fluctuations \cite{benet16}:
\begin{equation}\label{eq:h2surf}
   \begin{array}{ccc}
	 \langle |\zsq^2({\bf q})| \rangle & = &
	  \displaystyle{     \frac{k_BT}{A} \frac{g'' + \gamma_{wv} q^2}{[
		     \upsilon + g'' +
  \tilde\gamma_{iw} q^2 ][  g'' + \gamma_{wv} q^2 ] - [g''+\Delta g'' - \gamma_{iv} q^2]^2} } \\
	   & & \\
    \langle |\zqv^2({\bf q})| \rangle & = &
      \displaystyle{      \frac{k_BT}{A} \frac{\upsilon + g'' +
   \tilde\gamma_{iw} q^2}{[ \upsilon + g'' +
	\tilde\gamma_{iw} q^2 ][  g'' + \gamma_{wv} q^2 ] - [g''+\Delta g'' - \gamma_{iv} q^2]^2}  } \\
		& & \\
	 \langle \zsq({\bf q}) \zqv^*({\bf q}) \rangle & = &
	 \displaystyle{      \frac{k_BT}{A} \frac{g'' + \Delta g'' - \gamma_{iv} q^2}{[ \upsilon + g'' +
    \tilde\gamma_{iw} q^2 ][  g'' + \gamma_{wv} q^2 ] - [g''+\Delta g'' -
 \gamma_{iv} q^2]^2}  }
 \end{array}
\end{equation}
where $g''$ is the second derivative of the interface potential at
the equilibrium film thickness, $\Delta g''$ accounts for enhanced coupled
compression-expansion of the film thickness, and $\upsilon=q_z^2 u$.
In practice, in order to avoid under-determined fits to limited data,
we set $\Delta g''$ and $\gamma_{iv}$ to zero.  
This model for the spectrum of surface fluctuations has small
and large wave-vector regimes. At large wave-vectors, the spectrum
of fluctuations depends only on the stiffness and surface tension
coefficients, $\tilde\gamma_{iw}$ and $\gamma_{wv}$ respectively \cite{benet19}.
As in extended capillary wave Hamiltonians, 
these  are modeled as  
even polynomials of $q$ to order $4$. Once
$\tilde\gamma_{iw}$ and $\gamma_{wv}$ are known, we fit the remaining
parameters $g''$ and $\upsilon$ to match simultaneously
the low wave-vector regime of $q^2\langle \zsq(q)\zsq(q) \rangle$,
$q^2\langle \zqv(q)\zqv(q) \rangle$ and $q^2\langle \zsq(q)\zqv(q) \rangle$
as obtained from simulations. 
The model reproduces very accurately the stiffness coefficients at high
temperature. In the region were the DOF phase appears, it does however not
grasp the enhanced height fluctuations. However, quite generally, the
correlation functions may be expressed as a spectral series. The mean
field solution corresponds to the 'trivial' eigenmode which results for
the quadratic expansion of Hamiltonian. Higher order solutions yield
additional eigenmodes. For strictly periodical height profiles, the
spectrum of eigenmodes has a band structure. For the quasi-long range
observed in the DOF phase, we expect the spectrum will not have
continuous bands, but rather a region of close eigenvalues that are
separated from the mean field mode. Accordingly, the effect of
the inhomogeneity appears as an additive correction to the mean field
solution. This allows us to obtain meaningful fits to the regular part
of the spectrum merely by neglecting the strong enhancement at intermediate
length scales. 
The step free energy for the uncoupled
Sine Gordon model can be obtained from the parameters  $\tilde\gamma_{iw}$ and
$\upsilon$ as
$\beta_{iw}=(8/q_z^2)(\gamma_{iw}\upsilon)^{1/2}$ \cite{nozieres87,benet19}. 
Since we find that the premelting film thickness is unimodal, a step
on the \SqL surface will provoke a similar step on the \qLV surface,
with a step energy of $\beta_{wv}=(8/q_z^2)(\gamma_{wv}g'')^{1/2}$.
The cost of creating a step on the premelting film
is therefore the sum $\beta=\beta_{iw} + \beta_{wv}$,
which we use as an estimate for the step free energy of the 
ice/vapor interface in Fig.\ref{fig:beta}.

%Supplementary Text\\
%Figs. S1 to S3\\
%Tables S1 to S4\\
%References \textit{(4-10)}

% For your review copy (i.e., the file you initially send in for
% evaluation), you can use the {figure} environment and the
% \includegraphics command to stream your figures into the text, placing
% all figures at the end.  For the final, revised manuscript for
% acceptance and production, however, PostScript or other graphics
% should not be streamed into your compliled file.  Instead, set
% captions as simple paragraphs (with a \noindent tag), setting them
% off from the rest of the text with a \clearpage as shown  below, and
% submit figures as separate files according to the Art Department's
% instructions.

\section*{Acknowledgments}

We thank E. Lomba for helpful discussions and
J. L. F. Abascal for invaluable support. 

\subsection*{Funding}

We acknowledge funds from the  Spanish
Agencia Estatal de Investigaci\'on
under Grant No. FIS2017-89361-C3-2-P.
This project was made possible thanks to the use of
the Mare-Nostrum supercomputer and the technical support
provided by Barcelona Supercomputing Center
from the Spanish Network of Supercomputing (RES) under grants
QCM-2017-2-0008 and QCM-2017-3-0034.

\subsection*{Author Contributions}

P.L. performed programs and  simulations,  analyzed the data and discussed
results as part of his phD Thesis.  E.G.N. supervised and discussed results.
L.G.M. conceived the project,  interpreted results and wrote the paper.

\subsection*{Competing  interests}

The authors declare no competing financial interests.

%Here you should list the contents of your Supplementary Materials -- below is an example. 
%You should include a list of Supplementary figures, Tables, and any references that appear only in the SM. 
%Note that the reference numbering continues from the main text to the SM.
% In the example below, Refs. 4-10 were cited only in the SM.     
%\section*{Supplementary materials}

%Materials and Methods\\
%Supplementary Text\\
%Figs. S1 to S3\\
%Tables S1 to S4\\
%References \textit{(4-10)}

\clearpage

\subsection*{Supplementary Figures S1 to S8}

\begin{figure}[h!]
   \includegraphics[clip,scale=0.15]{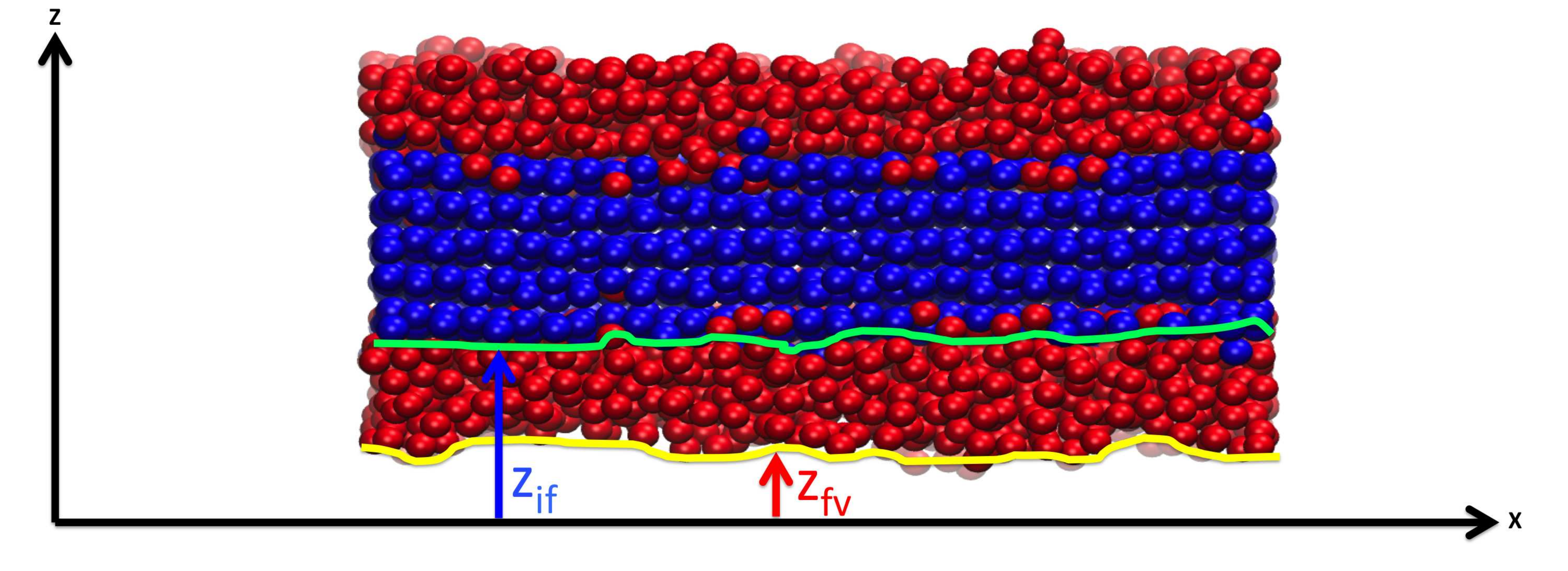}
   \caption*{\footnotesize {\bf Fig. S1: 
	   Characterization of the
	premelting layer.} Snapshot of a bulk ice slab  in equilibrium 
	with pure water vapor. After placing a slab of perfect ice in vacuum,  a premelting layer of 
	disordered water molecules
   is formed spontaneously in a few nanoseconds. Using a suitable order parameter it is possible to
   label liquid-like from solid-like molecules. The state of the premelting film may be described
   in terms of two different surfaces, $\zsq$ and $\zqv$, separating the film from bulk solid and
vapor phases, respectively.} 
%\label{fig:confi}
\end{figure}

%\vspace*{-0.1cm}

\begin{figure}[h!]
     \includegraphics[width=0.55\paperwidth,height=0.55\paperwidth,keepaspectratio,trim=0 0 0 0.8cm]{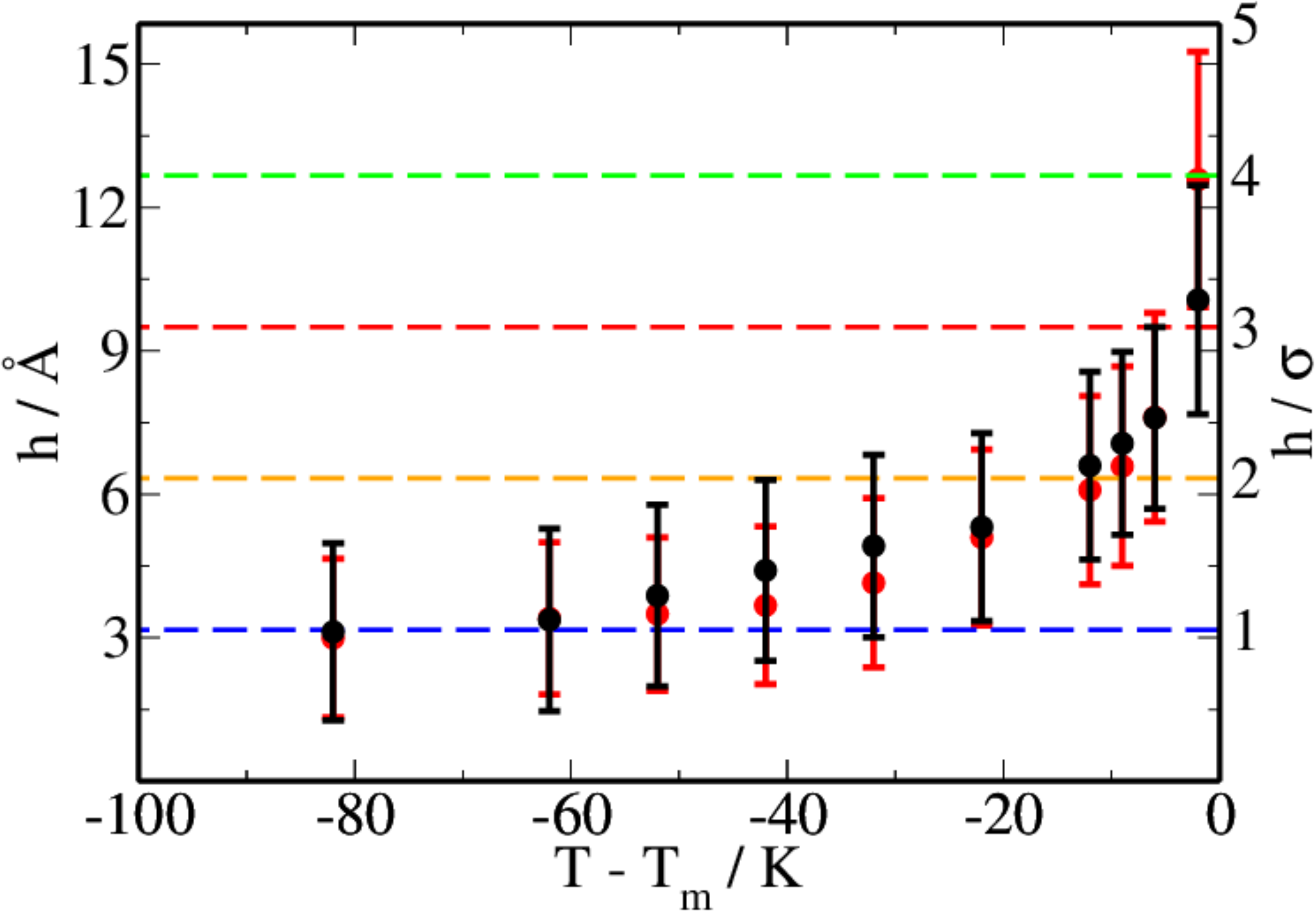}
     \caption*{\footnotesize {\bf Fig.S2: %1: 
	Premelting film thickness as a function of temperature for basal and
   prismatic faces.} Figure shows the thermally averaged film thickness $h$
   of the basal (black) and prismatic (red) planes. Dashed lines indicate
   multiples of the molecular diameter as measured in units of the Lennard-Jones
   $\sigma$ parameter. The thickness of the
prismatic plane remains only slightly below that of the basal plane up to
270~K, were the prismatic plane premelts by almost one full bilayer more} 
\end{figure}

\clearpage

\begin{figure}[h!]
   \includegraphics[clip,scale=0.20]{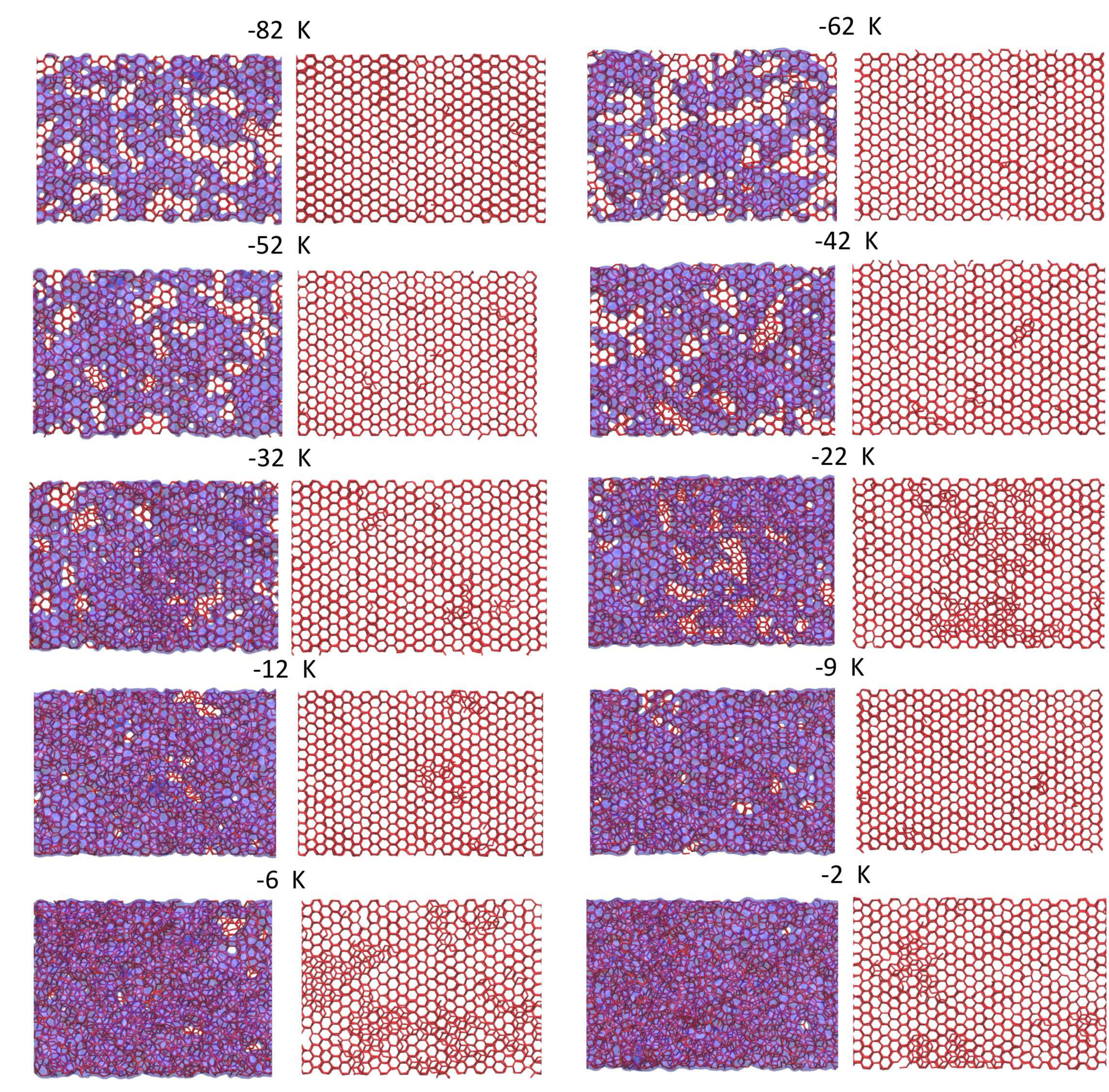}
   \caption*{\footnotesize {\bf  Fig.S3: %2: 
	  Evolution of surface structure with temperature on the 
     basal face}. 
     The first and third columns show the position of all atoms in the cluster
     of condensed molecules (solid-like or liquid-like),
     projected onto the x-y plane. The red wire-frame joins atoms separated by
     less than 3.5~{\AA}, which is the same criteria used to define first neighbours in the algorithm
     to search for solid clusters. Liquid-like atoms are
     further  coloured in violet.  The 
     typical hexagonal honeycomb  is clearly visible on patches not covered
     by premelted water-like molecules. Also notice how the premelted molecules often occupy 
     interstitial positions on the center of the hexagonal honeycomb.
     The second and third columns show only the positions of solid-like atoms,
     with liquid-like atoms left apart. As far as the position of oxygen
     atoms, the surface remains unreconstructed in all the
	  temperature domain. Patches of stacking disordered ice appear at
	  temperatures $\Delta T=-22$ and -2~K consistent with the melting of
	  complete solid bilayers.  
     }
%\label{fig:vmd_basal}
\end{figure}

\begin{figure}[h!]
   \includegraphics[clip,scale=0.20]{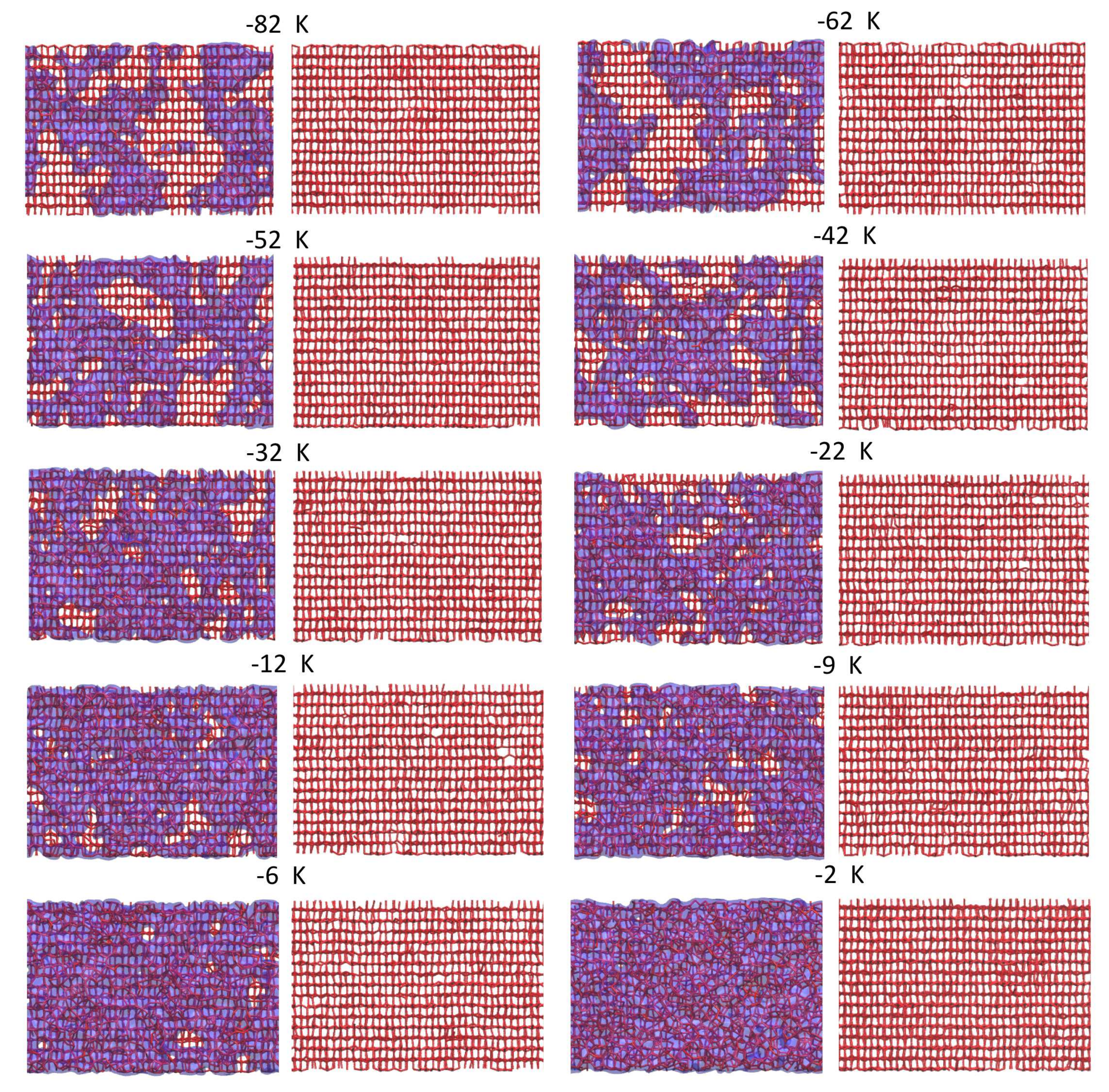}
   \caption*{\footnotesize {\bf  Fig.S4: %3: 
	  Evolution of surface structure with temperature on the 
     prismatic face}. 
     The first and third columns show the position of all atoms in the cluster
     of condensed molecules (solid-like or liquid-like)
     projected onto the x-y plane. The red wire-frame joins atoms separated by
     less than 3.5~{\AA}, which is the same criteria used to define first neighbours in the algorithm
     to search for solid clusters. Liquid-like atoms are
     further coloured in violet.  The 
     typical rectangular structure of prismatic faces  is clearly visible on patches not covered
     by premelted water-like molecules.  
     The second and third columns show only the positions of solid-like atoms,
     with liquid-like atoms left apart.  The oxygen framework
     remains unreconstructed in all the temperature domain.  }
%\label{fig:vmd_pi}
\end{figure}

\begin{figure}[h!]
   \includegraphics[clip,scale=0.22]{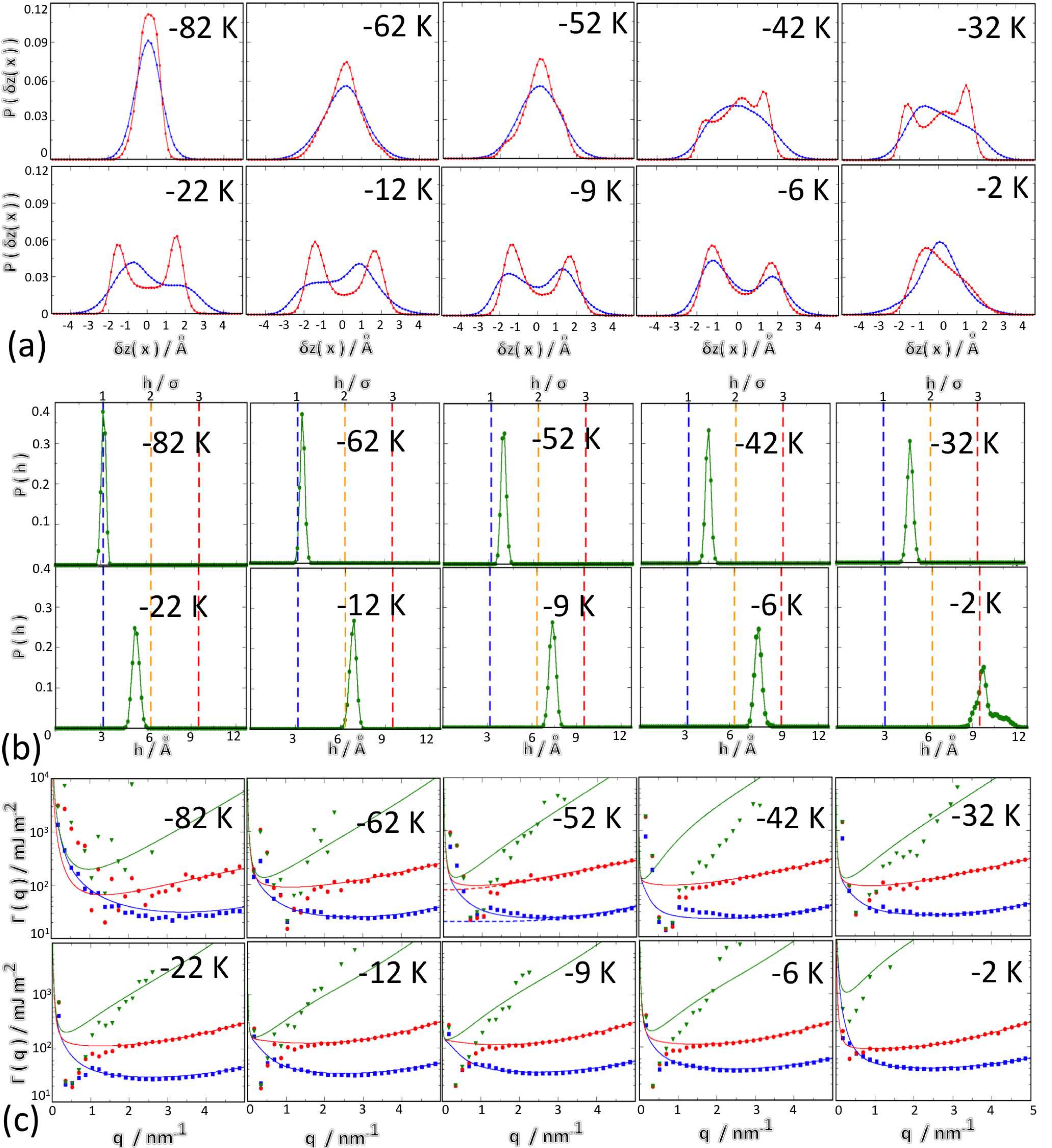}
     \caption*{\footnotesize 
	  {\bf Fig.S5: %4:  
		 Surface fluctuations on the basal face at
	 all studied temperatures}. a) Probability distribution
      of \SqL (blue) and \qLV (red) surface fluctuations, as measured by
	 the deviations of the interface position along $x$ about the
 average surface. 
 b) Probability distribution of the
          global premelting layer thickness, $h$, on the basal face for several
          temperatures as indicated in the color code. The vertical dashed lines
          indicate the location of multiples of the layer thickness in units of
          the molecular diameter ($\sigma$ parameter of the Lennard-Jones bead
	 in the TIP4P/Ice model). c) Spectrum of fluctuations on the basal face.
          The figure shows wave-vector dependent stiffness
          coefficients, as obtained from the inverse surface structure factor
          for \SqL correlations (blue), \qLV correlations (red) and crossed \SqL-\qLV
          correlations (green). Symbols are results from simulations. The full lines
          are a fit to the small wave-vector results to the SG+CW model, using
          as input results from the large wave-vector fit shown in dashed lines.
}
%\label{fig:ext_datas_basal}
\end{figure}

\begin{figure}[h!]
   \includegraphics[clip,scale=0.22]{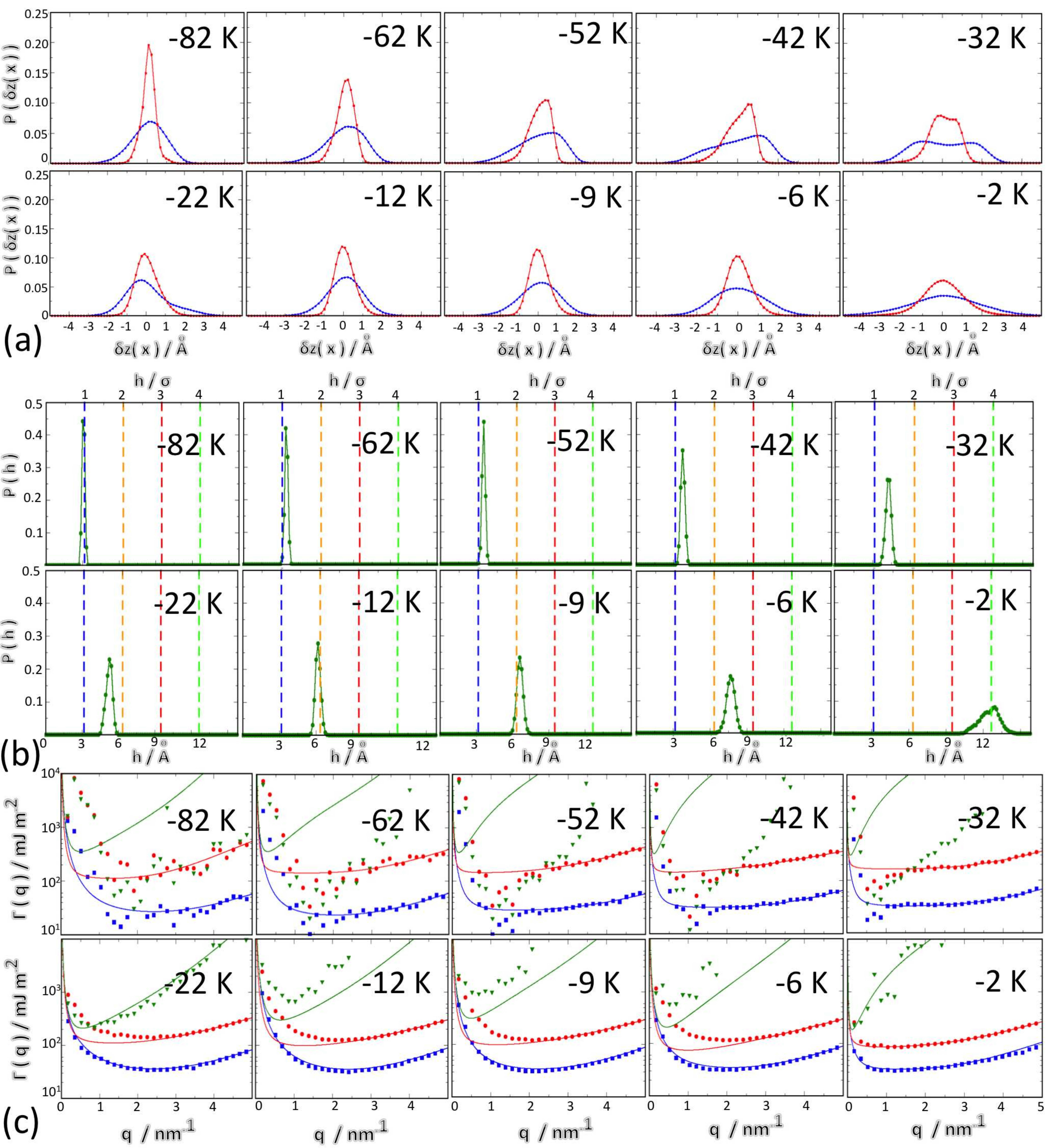}
     \caption*{\footnotesize 
	  {\bf Fig.S6: %5:  
		 Surface fluctuations on the prismatic face at
	 all studied temperatures}. a) Probability distribution
      of \SqL (blue) and \qLV (red) surface fluctuations, as measured by
	 the deviations of the interface position along $x$ about the
 average surface. 
 b) 
 Probability distribution of
 the local height fluctuations $\delta \z(\rpar)$. c)
          Probability distribution of the
          global premelting layer thickness, $h$, on the basal face for several
          temperatures as indicated in the color code. The vertical dashed lines
          indicate the location of multiples of the layer thickness in units of
          the molecular diameter ($\sigma$ parameter of the Lennard-Jones bead
	 in the TIP4P/Ice model). c) Spectrum of fluctuations on the basal face.
          The figure shows wave-vector dependent stiffness
          coefficients, as obtained from the inverse surface structure factor
          for \SqL correlations (blue), \qLV correlations (red) and crossed \SqL-\qLV
          correlations (green). Symbols are results from simulations. The full lines
          are a fit to the small wave-vector results to the SG+CW model, using
          as input results from the large wave-vector fit shown in dashed lines.
}
%\label{fig:ext_datas_pi}
\end{figure}

\begin{figure}[h!]
     \includegraphics[width=0.75\paperwidth,height=0.75\paperwidth,keepaspectratio]{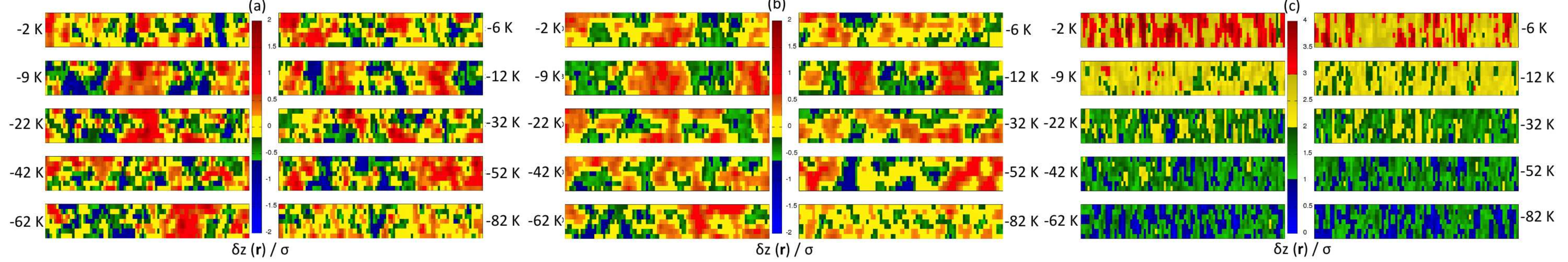}
     \caption*{\footnotesize
	  {\bf Fig.S7: %6: 
	     Surface plots of local height fluctuations
     for the basal face}. a) Local height fluctuations $\delta \zsq(\rpar)$  
  of the \SqL surface (see M1). b) Local height fluctuations $\delta \zqv(\rpar)$ of
  the \qLV surface (see M2). 
  At low and high temperatures, the correlation lengths are small and one finds
  alternated red and blue patches of small size. At intermediate temperatures
  there appear large red and blue patches indicative of the emergence of large
  correlations of preferred wave-length.  In this regime, comparison of a) and b) shows 
  that the \SqL and \qLV surfaces are also highly correlated. 
  c) Local fluctuations of film thickness $h(\rpar)$ (see M3). The correlation
  length of $\delta h(\rpar)$ remains small at all temperatures as is visible
  from the small size of alternating patches. The thickening of the films
  as temperature increases is apparent from the change in color code.
        }
%\label{fig:hxy_basal}
\end{figure}

\begin{figure}[h!]
     \includegraphics[width=0.75\paperwidth,height=0.75\paperwidth,keepaspectratio]{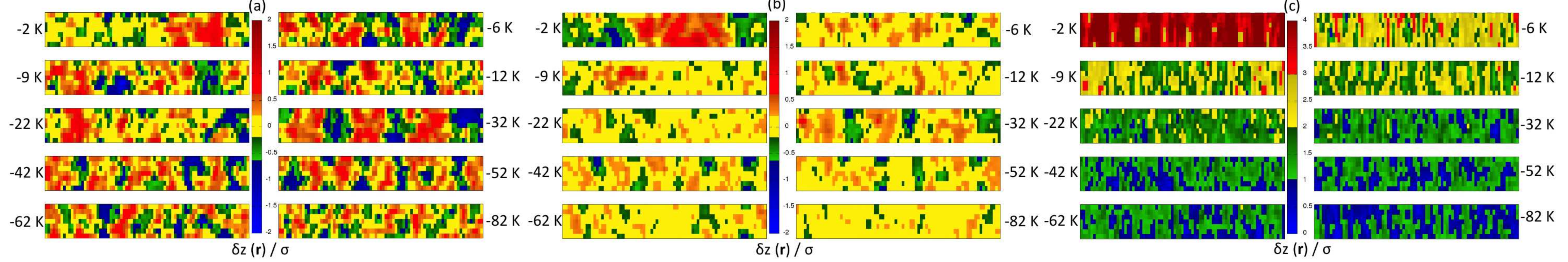}
     \caption*{\footnotesize
	  {\bf Fig.S8: %7: 
	     Surface plots of local height fluctuations
     for the prismatic face}. a) Local height fluctuations $\delta \zsq(\rpar)$  
  of the \SqL surface (see M4). At low temperatures the alternation of small blue and
  red domains is indicative of small correlation lengths. The size of the
  domains increases significantly at a DOF phase at T=240~K and then decrease
  again. At the highest temperature, T=270~K, a large correlated domain that
  spans most of the simulation cell is indicative of the approach to a
  roughening transition. b) Local height fluctuations $\delta \zqv(\rpar)$ of
  the \qLV surface (see M5). The temperature evolution of correlation lengths is similar
  to that of the \SqL, but compared to the basal surface, the amplitude of the
  fluctuations here is much smaller. c) Local fluctuations of film thickness $h(\rpar)$ (see M6). 
  As for the basal face, the correlation length of $\delta h(\rpar)$ remains small at all 
  temperatures as is visible from the small size of alternating patches. The thickening of the films
  as temperature increases is apparent from the change in color code.
        }
%\label{fig:hxy_pi}
\end{figure}

\clearpage

\subsection*{Supplementary Movies M1 to M6}

\begin{itemize}
   \item Movie M1: Local height fluctuations $\delta \zsq(\rpar)$  of the \SqL
	surface for the basal plane (see also Fig.S7).
	\item Movie M2: Local height fluctuations $\delta \zqv(\rpar)$  of the
	   \qLV surface for the basal plane (see also Fig.S7).
	\item Movie M3: Local fluctuations of film thickness $h(\rpar)$ for the
	   basal plane (see also Fig.S7).
\item Movie M4: Local height fluctuations $\delta \zsq(\rpar)$  of the \SqL
   surface for the  prismatic plane (see also Fig.S8).
\item Movie M5: Local height fluctuations $\delta \zqv(\rpar)$  of the \qLV 
   surface for the prismatic plane (see also Fig.S8)
   \item Movie M6:  Local fluctuations of film thickness $h(\rpar)$ for the
	prismatic plane (see also Fig.S8).
\end{itemize} 

\clearpage

\end{document}